\documentclass[prl,twocolumn,reprint,superscriptaddress]{revtex4-2}
\usepackage{amsmath}
\usepackage{times}
\usepackage{graphicx}
\usepackage{inputenc}
\usepackage{textcomp}  
\usepackage[hypertexnames=false]{hyperref}
\usepackage{comment}
\usepackage{color}
\usepackage{xcolor}

\newcommand {\ii}          {\mathrm{i}}
\newcommand {\bhzcoeff}[1] {\mathcal{#1}}
\newcommand {\ket}[1]      {\lvert#1\rangle}
\providecommand {\micro} {\textmu}

\DeclareMathOperator{\sgn}{sgn}

\DeclareGraphicsExtensions{.png,.pdf,.ai}
\renewcommand{\figurename}{\textsc{Figure}}
\renewcommand{\tablename}{\textsc{Table}}

\begin{document}

\title{Spectral asymmetry induces a re-entrant quantum Hall effect in a topological insulator}

\author{Li-Xian Wang} 
\affiliation{Institute for Topological Insulators, Am Hubland, 97074 W{\"u}rzburg, Germany}
\affiliation{Physikalisches Institut, Experimentelle Physik III, Universit{\"a}t W{\"u}rzburg, Am Hubland, 97074 W{\"u}rzburg, Germany}	
\author{Wouter Beugeling} 
\affiliation{Institute for Topological Insulators, Am Hubland, 97074 W{\"u}rzburg, Germany}
\affiliation{Physikalisches Institut, Experimentelle Physik III, Universit{\"a}t W{\"u}rzburg, Am Hubland, 97074 W{\"u}rzburg, Germany}
\author{Fabian Schmitt} 
\affiliation{Institute for Topological Insulators, Am Hubland, 97074 W{\"u}rzburg, Germany}
\affiliation{Physikalisches Institut, Experimentelle Physik III, Universit{\"a}t W{\"u}rzburg, Am Hubland, 97074 W{\"u}rzburg, Germany}
\author{Lukas Lunczer}
\affiliation{Institute for Topological Insulators, Am Hubland, 97074 W{\"u}rzburg, Germany}
\affiliation{Physikalisches Institut, Experimentelle Physik III, Universit{\"a}t W{\"u}rzburg, Am Hubland, 97074 W{\"u}rzburg, Germany} 
\author{Julian-Benedikt Mayer}
\affiliation{Institute for Theoretical Physics and Astrophysics (TP IV), Universit{\"a}t W{\"u}rzburg, Am Hubland, 97074 W{\"u}rzburg, Germany}
\author{Hartmut Buhmann}
\affiliation{Institute for Topological Insulators, Am Hubland, 97074 W{\"u}rzburg, Germany}
\affiliation{Physikalisches Institut, Experimentelle Physik III, Universit{\"a}t W{\"u}rzburg, Am Hubland, 97074 W{\"u}rzburg, Germany}
\author{Ewelina M. Hankiewicz}
\affiliation{Institute for Theoretical Physics and Astrophysics (TP IV), Universit{\"a}t W{\"u}rzburg, Am Hubland, 97074 W{\"u}rzburg, Germany}
\author{Laurens W. Molenkamp}
\affiliation{Institute for Topological Insulators, Am Hubland, 97074 W{\"u}rzburg, Germany}
\affiliation{Physikalisches Institut, Experimentelle Physik III, Universit{\"a}t W{\"u}rzburg, Am Hubland, 97074 W{\"u}rzburg, Germany}

\begin{abstract}
The band inversion of topological materials in three spatial dimensions is intimately connected to the parity anomaly of two-dimensional massless Dirac fermions. 
At finite magnetic fields, the parity anomaly reveals itself as a non-zero spectral asymmetry, i.e., a non-zero difference between the number of conduction and valence band Landau levels, due to the unpaired zero Landau level.
Here, we realize this two-dimensional Dirac physics at a single surface of the three-dimensional topological insulator (Hg,Mn)Te.
We observe an unconventional re-entrant quantum Hall effect that can be directly related to the occurrence of spectral asymmetry in a single topological surface state.
The effect should be observable in any topological insulator where the transport is dominated by a single Dirac surface state.

\end{abstract}
	
\date{\today}

\maketitle

\section*{Introduction}
\begin{figure}
    \centering
    \vspace{-5mm}
    \hspace*{-2mm}\includegraphics[width=90mm]{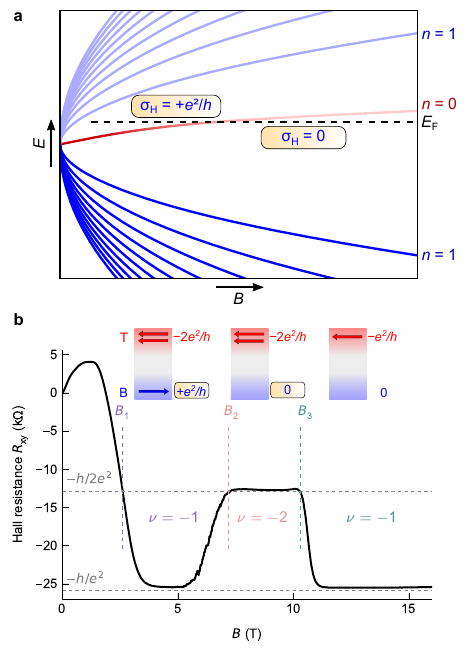}\\
    \vspace{-5mm}
    \caption{%
        \textbf{Reentrant quantum Hall effect in a three-dimensional topological insulator device due to spectral asymmetry.}
        \textbf{a}, Mechanism of spectral asymmetry. Upon increasing the magnetic field $B$, the band inversion is lost, and the zero Landau level moves from the valence band to the conduction band. The zero Landau level becomes unoccupied and its contribution of $e^2/h$ to the Hall conductance vanishes.
        \textbf{b}, Observation of the reentrant quantum Hall effect in the $p$-type regime. Upon increasing the magnetic field $B$, the Hall resistance $R_{xy}$ settles at a plateau at the quantized value of $-h/e^2$ ($\nu=-1$), followed by a transition to $-h/2e^2$ ($\nu=-2$) and a reentrance to $-h/e^2$. The schematics (insets) indicate the contributions of top and bottom surface to the Hall conductance. We identify the change in Hall conductance by $-e^2/h$ at the transition from $\nu=-1$ to $-2$ (at $B_2$) as the vanishing of the zero Landau level of the bottom surface state, due to spectral asymmetry. The transition from $\nu=-2$ to $-1$ (at $B_3$) is an ordinary $p$-type quantum Hall transition of the massive surface states at the top surface. This data has been taken for $V_\mathrm{TG}^{*}=-0.17$ V with bottom gate grounded. We mark the characteristic fields $B_1$, $B_2$ and $B_3$.
    }
    \label{fig:intro}
\end{figure}

Quantum field theory predicts that a single massless two-dimensional Dirac 
fermion exhibits the parity anomaly \cite{Redlich1984,semenoff1984condensed}. In a metal, the anomaly induces an ambiguity in the sign of the transverse conductivity in the limit of vanishing magnetic field \cite{semenoff1984condensed,fradkin1986physical}.
While the connection between the parity anomaly and topological surface states \cite{volkov1985two,Kusmartsev1985,KaneMele2005PRL95-14,konig2007quantum} has been hinted at extensively \cite{fradkin1986physical,haldane1988model}, one can argue that a true observation of the parity anomaly in a condensed matter system, i.e., in the limit of zero magnetic field, is still outstanding. Such an observation is challenging because topological surface states come in pairs and the anomaly can occur only within a single Dirac state \cite{Redlich1984,semenoff1984condensed,haldane1988model}. While in the quantum anomalous Hall effect one of the entangled surface states is removed, the effect occurs \cite{chang2013experimental,Mogi2022} only in ferromagnets. Thus, the hysteresis connected with ferromagnetism makes an observation of the anomaly at zero field very difficult.
In this work, we report on the experimental observation of a spectral asymmetry in the transport characteristics 
of the three-dimensional paramagnetic topological insulator (Hg,Mn)Te. 
The spectral asymmetry \cite{NiemiSemenoff1983}, a consequence of the parity anomaly \cite{Redlich1984,semenoff1984condensed,fradkin1986physical,MulliganBurnell2013}, induces an anomalous Hall response, i.e., an extra contribution of $e^2/h$ to the Hall conductance $\sigma_\mathrm{H}$ from the unpaired zero Landau level (see Figure~\ref{fig:intro}a), when the bulk bands are in inverted order \cite{bottcher2019survival,Bottcher2020}. 
In our magneto-transport experiments it manifests itself as a remarkable sequence of $\nu = -1, -2, -1$ quantum Hall plateaus for increasing magnetic fields. This observation requires the coexistence of topological Dirac surface states and massive (non-topological) surface states, where the $ \nu = -1 $ to $ \nu = -2 $ transition marks the crossing of an unpaired topological zero Landau level with the Fermi level, changing the Hall resistance $R_{xy}$ from $-h/e^2$ to $-h/2e^2$ (see\ Figure~\ref{fig:intro}b).
The parity anomaly should be accessible in any topological insulator device, provided that a single surface can be accessed in magneto-transport and that the material is of sufficiently high quality.

\section*{Results and discussion}

\begin{figure}
    \centering
    \vspace{-5mm}
    \hspace*{-2mm}\includegraphics[width=90mm]{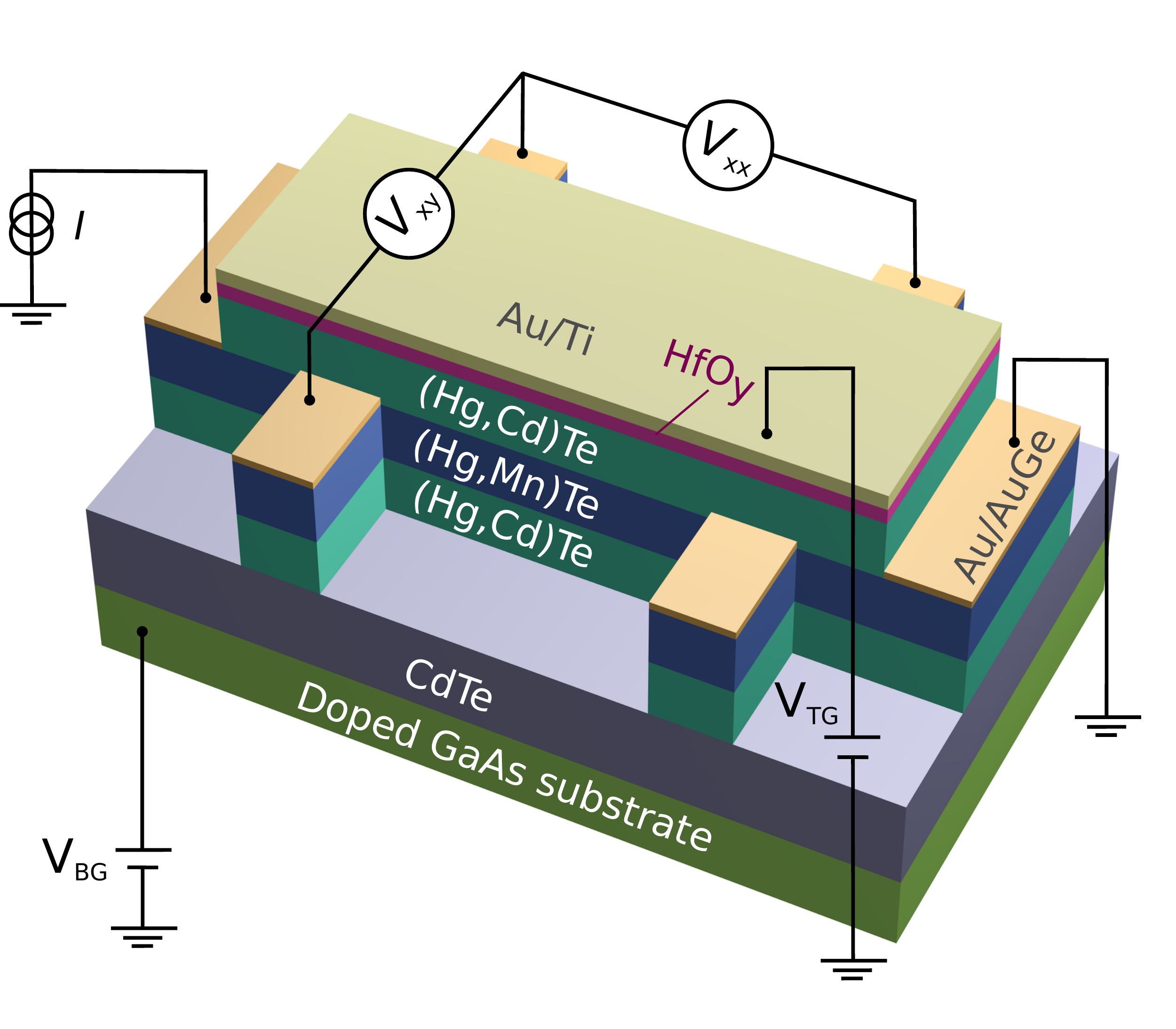}\\
    \vspace{-5mm}
    \caption{%
        \textbf{Structure of the Hall bar device.}
      Illustration of the layer structure (thickness not to scale) and the Hall measurement configuration. We indicate the (Hg,Mn)Te topological insulator layer and the (Hg,Cd)Te barriers. The doped GaAs substrate acts as bottom gate and is separated from the bottom barrier by a CdTe buffer. The Au/Ti top gate is separated by a thin insulating HfO$_y$ layer. The transport measurements are performed through the Au/AuGe Ohmic contacts.
    }
    \label{fig:device}
\end{figure}

\begin{figure}[t]
    \centering
    \hspace*{-2mm}\includegraphics[width=90mm]{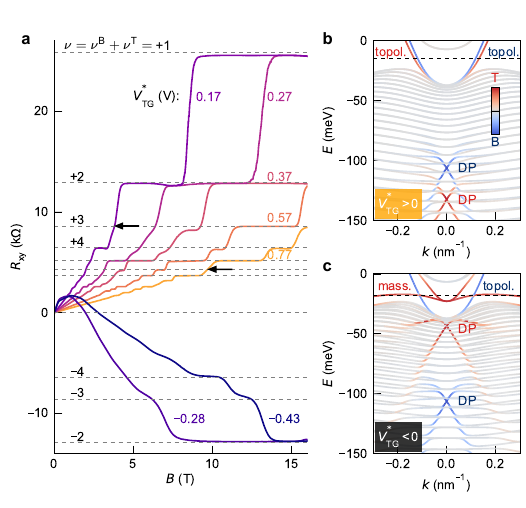}\\
    \caption{%
        \textbf{Ambipolar transport and quantum Hall effect in a 73 nm thick (Hg,Mn)Te layer.}
        \textbf{a}, Quantum Hall effect at various $V_\mathrm{TG}^{*}$, with bottom gate grounded. Two missing plateaus are indicated by arrows. The quantized values $R_{xy}=h/e^2\nu$ for integer $\nu$ are marked by dashed lines.
        The panels on the right show the band structure from an eight-orbital $k\cdot p$ calculation for two distinct scenarios: \textbf{b}, The $n$-type conducting regime $V_\mathrm{TG}^{*}>0$ and \textbf{c}, The $p$-type conducting regime $V_\mathrm{TG}^{*}<0$. We denote the Dirac point (DP), the topological surface states (``topol.''), and the massive surface states (``mass.'') in \textbf{b} and \textbf{c}. The dashed line indicates the Fermi energy. The color code indicates the wave function location: top surface (red), bottom surface (blue), or elsewhere (grey).
    }
    \label{fig:ambipolar_qhe}
\end{figure}

The transport data are obtained for a Mn-doped three-dimensional topological insulator device of HgTe. A 73~nm thick Hg$_{1-x}$Mn$_{x}$Te layer ($x=0.017$) is grown on a 4~\micro{}m thick CdTe buffer layer, sandwiched between thin (15 nm) barriers of non-inverted Hg$_{0.3}$Cd$_{0.7}$Te (see Figure~\ref{fig:device}). The lattice mismatch between (Hg,Mn)Te and CdTe provides the required tensile strain to turn the semimetallic (Hg,Mn)Te into a topological insulator with a bulk band gap of approximately $20$~meV \cite{brune2014dirac,leubner2016strain}. 
The Mn-doping of HgTe is isoelectrical and introduces paramagnetism which enhances the $g$-factor significantly \cite{Brandt1984,novik2005band,Shamim2020}, and predominantly increases the Landau level splitting already at low magnetic fields.
The transport results were obtained from a $600\times200$~\micro{}m$^2$ sized Hall bar equipped with top and bottom gates (see Figure~\ref{fig:device} and Experimental Section). Four additional samples have been investigated with Mn concentrations ranging between $1.1\%$ and $4.4\%$ and layer thicknesses between $64$ and $92$~nm (see Table~S1, Supporting Information). All samples exhibit the same magnetotransport feature discussed below (see Figure~S1, Supporting Information).

In Figure~\ref{fig:ambipolar_qhe}a, we show the low-temperature ($ T=120 $ mK) measurement of the Hall resistance $R_{xy}$, for various top gate voltages, ranging from $ -0.43 $~V to $ +0.77 $~V and grounded bottom gate ($V_\mathrm{BG} = 0$). 
For positive top gate voltages the conductance is purely n-type, while for negative gate voltages, two carrier types coexist: n-type carriers dominate at low and p-type carriers at high magnetic fields. The n-type carriers are characterized by low density and high mobility, while the p-type carriers come with a lower mobility but a large density \cite{mahler2021vps}.
Owing to the high quality of the samples a pronounced quantum Hall effect is observed for positive as well as for negative gate voltages. In order to compare measurements of different samples, the gate voltage is referenced to zero (i.e., we set the effective applied voltage $V_\mathrm{TG}^{*} = 0$~V) where the appearance of a two carrier type Hall effect becomes observable.
Missing steps in the Hall plateau sequence for positive gate voltages indicate the existence of two independent n-type carrier systems (see for example Figure~\ref{fig:ambipolar_qhe}a, $ \nu = 6 $ for $V_\mathrm{TG}^{*} = 0.77$~V, arrow). Referring to Ref.~\cite{brune2014dirac}, we identify them as topological surface states related to the top and bottom surfaces.

The appearance of p-type carriers for negative gate voltages is related to gate voltage induced massive surface states, as is typical for any topological insulator. The gate voltage pulls the non-topological massive surface states (also known as massive Volkov-Pankratov states \cite{volkov1985two}) out of the bulk valence band \cite{mahler2021vps}. The corresponding band structure calculation is given in Figures~\ref{fig:ambipolar_qhe}b and \ref{fig:ambipolar_qhe}c for positive and negative top gate voltages, respectively. This result confirms the observation that for positive gate voltages two massless topological surface states dominate the transport properties (Figure~\ref{fig:ambipolar_qhe}b) while for negative gate voltages an additional p-type massive surface state, located at the top surface contributes to transport (Figure~\ref{fig:ambipolar_qhe}c). Note that the high density of states of the valence band (van Hove singularity) pins the Fermi level. For large negative gate voltages, the massive surface state is pulled out of the bulk (cf.\ Figure~\ref{fig:ambipolar_qhe}c), which effectively empties the topological surface state at the top surface, while negative charge still exists in the bottom topological surface state.

\begin{figure}
    \centering
    \hspace*{-2mm}\includegraphics[width=90mm]{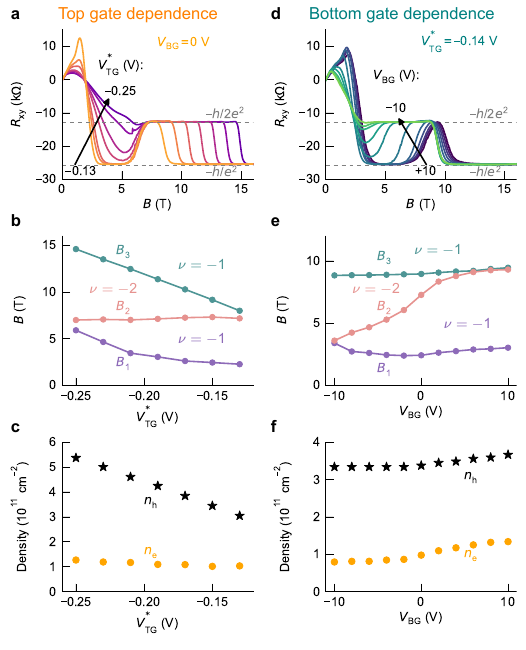}\\
    \caption{%
        \textbf{Re-entrance of the $\nu=-1$ quantum Hall plateau and its dependence on top and bottom gate voltages and carrier densities.}
        Top gate dependence \textbf{a} to \textbf{c} (the bottom gate is grounded):
        \textbf{a}, $R_{xy}$ as a function of magnetic field $B$ at various top gate voltages from $V_\mathrm{TG}^{*}=-0.13$ V to $-0.25$ V. 
        \textbf{b}, Characteristic fields extracted from \textbf{a}. The quantum Hall filling factor $\nu$ between characteristic fields is indicated by the circled numbers.
        \textbf{c}, Electron $n_\mathrm{e}$ and hole density $n_\mathrm{h}$.
        \textbf{d}--\textbf{f}, Like \textbf{a}--\textbf{c}, but for bottom gate dependence, at constant top gate voltage $V_\mathrm{TG}^{*}=-0.14$ V.
    }
    \label{fig:reentrance}
\end{figure}

The interplay between massive and topological surface states becomes most strikingly visible in Hall resistance $ R_{xy} $ for top gate voltages between $ V_\mathrm{TG}^* = -0.13 $ and $ -0.25 $ V. Figure~\ref{fig:intro}b shows the situation for $ V_\mathrm{TG}^* = -0.17$~V. Slightly above $ B = 1 $ T the initial positive slope in $ R_{xy} $ becomes negative and a clear $ \nu = -1 $ quantum Hall plateau develops for $ B > 4 $ T. Strikingly, for $ B > 5 $~T, $R_{xy}$ decreases again and a $ \nu = -2 $ quantum Hall plateau develops which lasts up to $10$~T before it reenters the $ \nu = -1 $ state. 
For the following discussion we distinguish the two $ \nu = - 1 $ states as low-field and high-field and define three magnetic fields, $ B_1$, $B_2$, and $B_3$, which mark the crossings of $ R_{xy} $ with the $ \nu = -2 $ line ($ R_{xy}(B_i) = -h/2e^2 $; cf.\ Figure~\ref{fig:intro}b).

To investigate the origin of the re-entrant behavior, we make use of the fact that our device is equipped with separate top and bottom gates. Importantly, due to screening, the electrostatic gates influence mainly those surface states which are in closest proximity. Figure~\ref{fig:reentrance} shows the variation of $ R_{xy} $ as a function of either top or bottom gate voltage (left and right column, respectively), while the other gate is kept at constant voltage.
For a variation of the top gate voltage from $-0.13$ to $-0.25$~V (Figure~\ref{fig:reentrance}a), we observe a gradual vanishing of the low-field $ \nu=-1 $ plateau, while $ B_2 $ remains almost unchanged. At the same time, the onset of the high-field $-1$ plateau ($ B_3 $) shifts to higher magnetic fields. The deduced characteristic magnetic fields, $ B_1$,  $ B_2 $, and $ B_3 $, are  shown in Figure~\ref{fig:reentrance}b. Analyzing the corresponding carrier densities, we find that the n-type carrier density (initial Hall slope) remains effectively constant while the p-type density increases monotonically with increasingly negative top gate voltage (Figure~\ref{fig:reentrance}c). 
The behaviour is consistent with the idea that the top gate voltage mainly affects the top surface states, increasing the p-type carrier density $n_\mathrm{h}$ of the massive surface state while the carrier density $n_\mathrm{e}$ of the n-type top topological surface state is pinned by the van Hove singularity.
The situation is opposite for a variation of the bottom gate voltage (Figures~\ref{fig:reentrance}d--f). Due to the different dielectric constants and the thickness of the insulating layer, the efficiency of the bottom gate is approximately two order of magnitude less than for the top gate. We find that the low-field $ \nu = -1 $ plateau is largest for $V_\mathrm{BG} = +10$~V and gradually vanishes in the range down to $V_\mathrm{BG} = -10$~V. In this case, $ B_1 $ and $ B_3 $ remain constant while $ B_2 $ changes strongly, and correspondingly, the n-type density $n_\mathrm{e}$ exhibits a stronger decrease than the p-type density $n_\mathrm{h}$.

From the above observations, we conclude that the low-field $\nu=-1$ plateau is the result of the interplay of quantized Hall conductances for top and bottom surface states. The combined total quantum Hall filling factor, which is observed in the measurement, can be divided into independent contributions from both surfaces:  $\nu=\nu^\mathrm{B}+\nu^\mathrm{T}$. The component $\nu^\mathrm{B}$ is entirely related to the n-type carrier density of the topological bottom surface states. The component $\nu^\mathrm{T}$ is controlled by the carrier density of the top surface, which is dominated by the massive p-type surface state. Its density is correlated with the magnetic field values for $ B_1 $ and $ B_3 $. 
At $V_\mathrm{BG}=-10$~V, the bottom surface state is partially depleted, thus the low-field $\nu=-1$ is absent in the transport experiment (see also Figure~S3, Supporting Information). However, the bottom surface state is most populated at $V_\mathrm{BG}=+10$~V and the low-field $\nu=-1$ state has the widest range in magnetic field.
We may relate the transition of $ \nu = -1 $ to $ -2 $ and back to $ -1 $ to a sequential change of the filling factor for top and bottom surface state as indicated in Figure~\ref{fig:intro}b: For the low-field $\nu=-1$ plateau, $\nu^\mathrm{B}+\nu^\mathrm{T}= 1 + (-2)$, the intermediate-field plateau has filling factors $\nu^\mathrm{B}+\nu^\mathrm{T}= 0 + (-2)$, and the high-field plateau has $\nu^\mathrm{B}+\nu^\mathrm{T}= 0 + (-1)$.  
Thus, the open question is to identify the transition of the topological surface state from filling factor $ \nu^\mathrm{B} = 1 $ to $ \nu^\mathrm{B} = 0 $, characterized by $ B_2 $ and controlled by the carrier density of the bottom surface state, as coming from spectral asymmetry.

\begin{figure}
    \centering
    \hspace*{-2mm}\includegraphics[width=90mm]{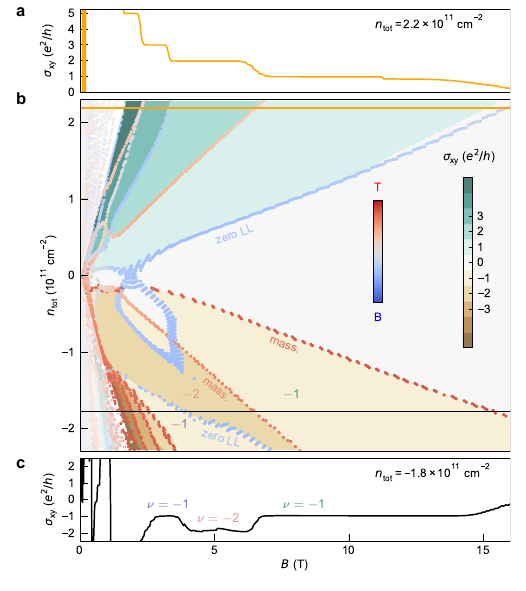}\\
    \caption{%
        \textbf{Landau level fan chart reproduced by an eight-orbital $k\cdot p$ calculation.}
        \textbf{b}, The color code in the background is associated with the calculated Hall conductivity $\sigma_{xy}$ in unit of $e^2/h$. The Landau levels are indicated by the dotted dispersions and their color indicates the wave function location: top surface (red), bottom surface (blue), or elsewhere (grey).
        The vertical axis represents the total density $n_\mathrm{tot}=n_\mathrm{e}-n_\mathrm{h}$, calculated from an assumed Hartree potential $U_\mathrm{H}$ (see Supplementary Note 4 and Figure~S4 for a detailed explanation). The re-entrant $\nu=-1$ quantum Hall plateau emerges for $n_\mathrm{tot} < -1.3\times10^{11}\ \mathrm{cm^{-2}}$.
        The most relevant Landau levels in this re-entrance are highlighted: ``zero LL'' for the zero Landau level from the topological surface state and ``mass.'' for the Landau levels from the massive surface state (at the bottom and top surface, respectively). The ``zero LL'' appears at multiple densities because the Fermi level depends non-monotonically on $n_\mathrm{tot}$ due to pinning to the massive surface state (see Supplementary Note 4).
        \textbf{a}, Calculated Hall conductivity $\sigma_{xy}$ at total density $n_\mathrm{tot}=2.2\times10^{11}\ \mathrm{cm^{-2}}$ (yellow line in \textbf{b}) and
        \textbf{c}, Calculated Hall conductivity $\sigma_{xy}$ at total density $n_\mathrm{tot}=-1.8\times10^{11}\ \mathrm{cm^{-2}}$ (black line in \textbf{b}). For Landau fans at these densities, refer to Figure~S4 (Supporting Information).
    }
    \label{fig:llfan}
\end{figure}

In order to substantiate our assignment of the origin of the re-entrant quantum Hall effect, we perform Landau level calculations using an eight-orbital $k\cdot p$ method. We present Hall conductivity $\sigma_{xy}$ as a function of magnetic field in the range $ B = 0$ to $16$~T and of the total density $n_\mathrm{tot}$ inferred from the measurement, by varying the Hartree potential $U_\mathrm{H}$ (see Figure~\ref{fig:llfan}b and more details in Supplementary Note 4). This figure additionally has the color coded information on the location  of the carrier system responsible for the indicated Landau level transition (red, top (T) and blue, bottom (B) surfaces, respectively).
This method allows us to examine the occupation of each Landau level at a constant total density (gate voltage or Hartree potential) and thus to extract the corresponding filling factors $\nu=\nu^\mathrm{B}+\nu^\mathrm{T}$.
Two exemplary Hall traces are given in Figures~\ref{fig:llfan}a and \ref{fig:llfan}c for total carrier densities of $ 2.2 \times 10^{11}$ and $ -1.8 \times 10^{11}\ \mathrm{cm^{-2}}$, corresponding to $ V^*_\mathrm{TG} = 0.1$~V and $ V^*_\mathrm{TG} = -0.2$~V, respectively and indicated by horizontal solid lines in Figure~\ref{fig:llfan}b.
For positive gate voltages (i.e., $n_\mathrm{tot}=2.2\times10^{11}\ \mathrm{cm}^{-2}$, yellow line), a monotonic integer sequence of quantum Hall plateaus appears (Figure~\ref{fig:llfan}a and Figure~S4b,c, Supporting Information), as Landau levels from the top and bottom surface states are depopulated. Note that the $\nu=4$ plateau is missing due to the crossing of Landau levels from top and bottom surface states, similar to the phenomenology of the experiments.
For $n_\mathrm{tot}<-1.3\times10^{11}\ \mathrm{cm^{-2}}$ (Figure~\ref{fig:llfan}c and Figure~S4d,e, Supporting Information) we clearly observe the re-entrant sequence, $\nu = -1$, $-2$, $-1$, comparable with the presented experiment.
By examining the wave function profile ($|\psi(z)|^2$, where $z$ is the growth direction) and the orbital character of the Landau levels (see Figure~S5, Supporting Information), we establish their origin:
the low-field $\nu=-1$ state is a combination of $\nu^\mathrm{B}=1$ from the topological (bottom) surface state and $\nu^\mathrm{T}=-2$ from the massive surface state at the top surface (see Figure~\ref{fig:intro}b). At $B_2$, the bottom surface state transitions to $\nu^\mathrm{B}=0$ while $\nu^\mathrm{T}=-2$ remains unchanged.

The transition of $\nu^\mathrm{B}=1$ to $\nu^\mathrm{B}=0$ represents a single Dirac fermion state in two spatial dimensions and is a signature of the spectral asymmetry related to the parity anomaly known from quantum electrodynamics.
This transition occurs when the zero Landau level of the topological surface state crosses the Fermi level in the vicinity of $B_2$, reducing the Hall conductance by $e^2/h$, i.e., the total filling factor, $\nu=\nu^\mathrm{B}+\nu^\mathrm{T}$, changes from $ \nu = 1 +(-2) $ to $ \nu = 0 +(-2) $. At this position the zero Landau level changes from the valence band to the conduction band regime which causes a change of the spectral asymmetry, i.e., the scenario illustrated by Figure~\ref{fig:intro}.

To make this statement more precise, we describe the bottom surface state by the effective Dirac-like Hamiltonian in the basis $\{ \ket{\text{B};\uparrow}, \ket{\text{B};\downarrow} \}$, given by
\begin{align}
    H_\text{B}
    &= \bhzcoeff{C} \sigma_0 + m_k \sigma_z - \bhzcoeff{A}(k_y\sigma_x - k_x \sigma_y)\\ \nonumber
    &=\begin{pmatrix}
        \bhzcoeff{C} + m_k & -i\bhzcoeff{A}k_-\\
        i\bhzcoeff{A}k_+ & \bhzcoeff{C} - m_k
    \end{pmatrix},
\end{align}
where $m_k = \bhzcoeff{M} - \bhzcoeff{B} |k|^2$ is an effective mass term which is only present in finite magnetic fields, $\sigma_0$ is the $2\times 2$ identity matrix, $\sigma_{x,y,z}$ are the Pauli matrices in the spin basis, $\mathbf{k}=(k_x,k_y)$ is the momentum, and $k_\pm=k_x\pm \ii k_y$.
The contribution of the spectral asymmetry to the Hall conductance is given by (see Supplementary Note 5 and Ref.~\cite{bottcher2019survival})
\begin{align}
 \label{eq:sxyBSSmain}
 \sigma_{xy}^{\mathrm{B}}
 =  - \frac{e^2}{2h} \left[\sgn(\bhzcoeff{B}) + \sgn\left(\bhzcoeff{M} - \frac{\bhzcoeff{B}}{l_B^2}\right)\right],
\end{align}
where $l_B=\sqrt{\hbar/eB}$ is the magnetic length corresponding to the magnetic field $B$. We emphasize that the coefficients $\bhzcoeff{M}$ and $\bhzcoeff{B}$ also depend on the magnetic field. We assume $\bhzcoeff{B}<0$.
The spectral asymmetry manifests itself as a jump of the conductance $\sigma_{xy}^{\mathrm{B}}$ from $e^2/h$ ($\nu^\mathrm{B}=1$) to zero ($\nu^\mathrm{B}=0$), where the effective mass term, the second term in Equation~\eqref{eq:sxyBSSmain}, changes from negative to positive. This jump occurs exactly where the zero Landau level crosses from the valence to the conduction band (see Figure~\ref{fig:intro}a).
The sign change of the mass gap indicates the transition from inverted to normal band ordering, a fundamental feature of topological materials. 

\section*{Conclusion}

In conclusion, we demonstrate the observation of a single two-dimensional Dirac system, realized by separate control of the carrier densities on both surfaces of a three-dimensional topological insulator. 
The re-entrant sequence of quantum Hall plateaus allows us to unambiguously identify a single two-dimensional Dirac fermion. 
At this transition from $\nu = -1$ (low-field) to $ \nu = -2$ the contribution of $e^2/h$ from the bottom surface state vanishes, which is direct evidence for the presence of a spectral asymmetry. Importantly, the contribution from spectral asymmetry persists in the limit of zero magnetic field, $B\to0$.
This observation is thus a robust signature for a solid-state analogue to the parity anomaly.


\section*{Experimental Section}

\noindent\textbf{Sample description}\\
The samples have been grown on a Si-doped GaAs substrate which serves as a bottom gate. The transport results have been obtained from $600~\text{\textmu m} \times 200~\text{\textmu m}$ sized Hall bars. The samples are equipped with Au top gates ($100$ nm thick) separated from the rest of the structure by a $15$~nm thick hafnium oxide (HfO$_y$) insulating layer and a $5$ nm Ti layer. The ohmic contacts are made of 50~nm AuGe and 50~nm Au.
The sample for which the results are shown in the main text is labelled S1. We provide information on four further (Hg,Mn)Te samples (labelled S2--S5) in Table~S1 (Supporting Information). All the devices exhibit a plateau transition from the low-field $\nu=-1$ to $\nu=-2$ (Figure~S1, Supporting Information), similar to the phenomenology discussed in the main text.

\medskip
\noindent\textbf{Transport measurement}\\
The magneto-transport measurements have been performed with low-frequency lock-in technique in ${}^3$He-${}^4$He dilution refrigerator with a base temperature of 120~mK, unless specified otherwise.

\section*{Supporting information}
Supporting information is available with this manuscript.

\section*{Acknowledgements}
This work was supported by
the Deutsche Forschungsgemeinschaft (DFG, German Research Foundation) through the Leibniz Program (L.W.M.) and through the projects SFB 1170 (Project ID 258499086; L.W.M., H.B., E.M.H.) and SPP 1666 (Project ID 220179758; L.W.M., H.B.),
by the EU ERC-AdG program (Project 4-TOPS; L.W.M.),
by the W\"urzburg-Dresden Cluster of Excellence on Complexity and Topology in Quantum Matter (EXC 2147, Project ID 39085490; L.W.M., H.B., E.M.H.),
and by the Free State of Bavaria for the Institute for Topological Insulators (L.W.M.).

\section*{Author contributions}
L.X.W., L.W.M., and H.B. conceived the experiments.
L.L. grew the sample.
F.S. fabricated the device.
L.X.W. performed the magnetotransport measurements and the data analysis.
W.B., J.B.M., and E.M.H. performed the theoretical modeling.
W.B. developed the code used for the numerical analysis.
L.X.W. and W.B. wrote the manuscript with the input from all authors.

\section*{Competing interests}
The authors declare no competing interests.

\section*{Keywords}
Topological insulators,
narrow-gap semiconductors,
magnetotransport,
quantum Hall effect,
quantum anomalies


%

\onecolumngrid
\clearpage

\begin{center}
{\bf Supporting Information for:\\[.25em]}
{\large\bf Spectral asymmetry induces a re-entrant quantum Hall effect in a topological insulator\\}
{}
\vspace{1em}
{Li-Xian Wang, Wouter Beugeling, Fabian Schmitt, Lukas Lunczer, Julian-Benedikt Mayer, Hartmut Buhmann,\\Ewelina M.\ Hankiewicz, and Laurens W.\ Molenkamp}

\end{center}

\renewcommand{\thetable}{S\arabic{table}}%
\setcounter{figure}{0}
\renewcommand{\thefigure}{S\arabic{figure}}%
\setcounter{equation}{0}
\renewcommand{\theequation}{S\arabic{equation}}%

\renewcommand{\figurename}{\textsc{Figure}}
\renewcommand{\tablename}{\textsc{Table}}

\section{Supplementary Note 1: Additional samples}

\begin{table*}[h]
\begin{tabular}{p{40mm}p{40mm}p{40mm}}
\hline
\textbf{Device label} &\textbf{Mn concentration $x$} & \textbf{Layer thickness $d$ (nm)}\\ [1ex] 
\hline
S1 (presented)   & 0.017   & 73\\
S2  & 0.011   &81\\
S3 & 0.031 & 64\\
S4 & 0.017 & 74\\
S5 & 0.044 & 92\\ [1ex] 
\hline
\end{tabular}
\caption{%
    {\bf Measured Hg$_{1-x}$Mn$_{x}$Te samples with varying thickness and Mn concentration.} The measured samples all exhibit re-entrant quantum Hall effect at low temperatures, see Figure~\ref{fig:ext_moresamples}. The presented sample is labelled ``S1'', and the rest are labelled from ``S2'' to ``S5'', respectively.}
\label{table:1}
\end{table*}

\begin{figure*}[h]
  \includegraphics[width=110mm]{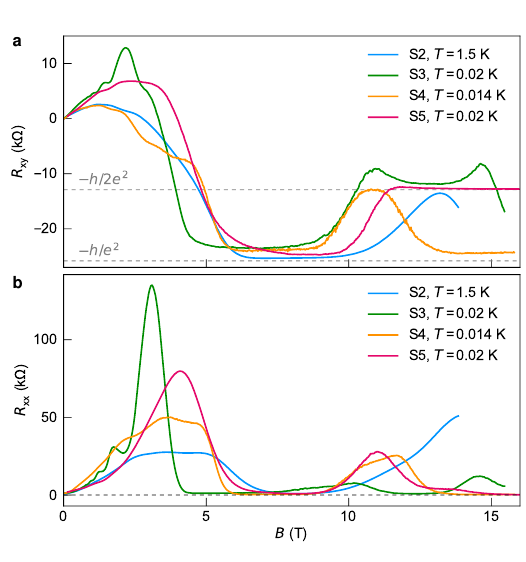}\\
  \caption{%
  		\textbf{Four further samples with varying Mn concentration and thickness.}
		\textbf{a}, Hall resistance $R_{xy}$ and
		\textbf{b}, Longitudinal resistance $R_{xx}$ as function of magnetic field $B$. The measurement temperature $T$ of each curve is also labelled. For detailed information for each sample, refer to Table~\ref{table:1}.
    }
  \label{fig:ext_moresamples}
\end{figure*}

\clearpage

\section{Supplementary Note 2: Extraction of electron and hole densities}

\begin{figure*}[h]
  \includegraphics[width=90mm]{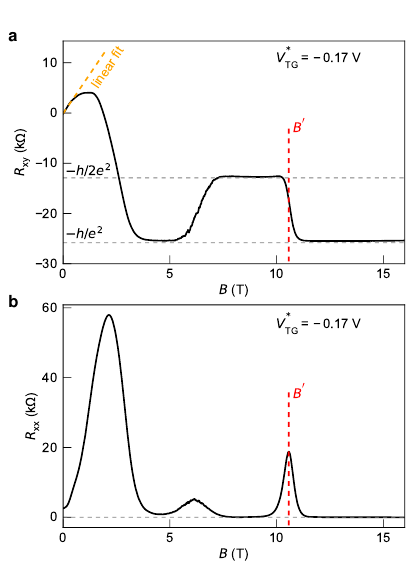}\\
  \caption{%
    \noindent {\bf Extraction of electron and hole densities.}
    \textbf{a}, Hall resistance $R_{xy}$ for $V_\mathrm{TG}^{*}=-0.17$~V, replotted from Figure~1b (main text). The slope of the linear fit at small magnetic fields is used to determine the n-type density $n_\mathrm{e}$. We indicate the field $B'$ at the transition between the $\nu=-2$ and $\nu=-1$ plateaus at high magnetic fields, from which we extract the p-type density $n_\mathrm{h}$.
    \textbf{b}, The value $B'$ is defined by the maximum of $R_{xx}$ corresponding to this transition.
    }
  \label{fig:ext_density}
\end{figure*}

The magneto-transport data of Figures~1b and 4a (main text) shows a clear two carrier behavior for negative top gate voltages. The n-type carriers with high mobility dominate the low field transport, and the p-type carriers dominate at high fields due to a lower mobility. The slope of the Hall resistance at low magnetic fields (see Figure~\ref{fig:ext_density}a) can be related to the n-type carrier density $n_\mathrm{e}$ by $n_\mathrm{e}=1/ (e\,dR_{xy}/dB)$, where $e$ is the elementary charge.
In the p-type dominated regime, at high magnetic fields, the quantum Hall effect is already well developed. We extract the field $B'$ of the transition from the $\nu=-2$ to the $\nu=-1$ plateau at high fields, which comes purely from the p-type carriers. For increased accuracy, we determine $B'$ from the corresponding maximum of $R_{xx}$, see Figure~\ref{fig:ext_density}b. Thus, we find the p-type density as $n_\mathrm{h}=-\overline{\nu} B'e/h$ with $\overline{\nu}=-\frac{3}{2}$.

\clearpage

\section{Supplementary Note 3: Low field quantum Hall plateau}

\begin{figure*}[h]
   \centering
   \includegraphics[width=110mm]{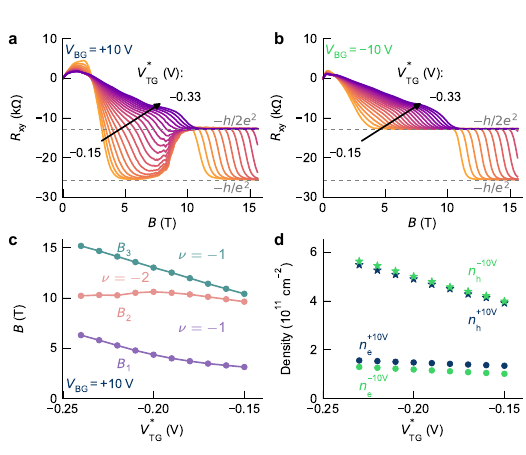}\\
   \caption{%
       \textbf{Low-field $\nu=-1$ quantum Hall plateau and its relation to the bottom surface state.}
       \textbf{a}, Hall resistance $R_{xy}$ as a function of magnetic field from $V^{*}_\mathrm{TG}=-0.15$ V to $-0.33$ V at $V_\mathrm{BG}=+10$ V.
       \textbf{b}, $R_{xy}$ as a function of magnetic field from $V^{*}_\mathrm{TG}=-0.15$ V to $-0.33$ V at $V_\mathrm{BG}=-10$ V. $R_{xy}=-h/e^2,-2h/e^2$ are indicated by dashed lines.
       \textbf{c}, Characteristic fields extracted from \textbf{a} for $V^{*}_\mathrm{TG}=-0.24$ V to $-0.15$ V. The quantum Hall filling factor $\nu$ between characteristic fields is indicated by the number in a circle.
       \textbf{d}, Electron density for $V_\mathrm{BG}=+10$ V ($n_\mathrm{e}^{+10V}$) and $-10$ V ($n_\mathrm{e}^{-10V}$) and hole density for $V_\mathrm{BG}=+10$ V ($n_\mathrm{h}^{+10V}$) and $-10$ V ($n_\mathrm{h}^{-10V}$) for $V^{*}_\mathrm{TG}=-0.24$ V to $-0.15$ V.
   }
   \label{fig:ext_qhe_cmpr}
\end{figure*}

\clearpage
\section{Supplementary Note 4: Calculation of experimentally relevant Landau fan}

In this section, we introduce the method used to calculate the Landau level fan chart displayed in the main text (Figure~5). Conventional fans charts display the energy dependence of each Landau level. However, in actual magneto-transport experiments, one obtains Hall conductance as a function of gate voltage rather than energy.
Changing the gate voltage not only varies the total carrier density $n_\mathrm{tot}=n_\mathrm{e}-n_\mathrm{h}$, but also the electrostatic potential. 

\begin{figure*}[h]
  \includegraphics[width=150mm]{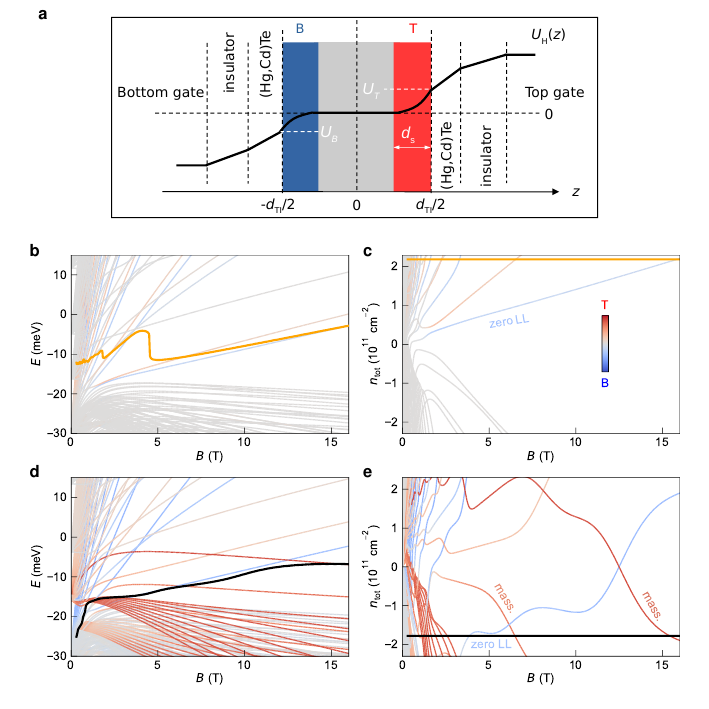}\\  
  \caption{%
  		\textbf{Landau level calculations for positive and negative top gate voltage.}
  		\textbf{a}, Sketch of Hartree potential $U_\mathrm{H}(z)$ as a function of position $z$ in crystal growth direction. $U_\mathrm{T}$ and $U_\mathrm{B}$ indicate the Hartree potential at the top and bottom interfaces, respectively. $d_\mathrm{s}$ stands for thickness of surface-state region (labelled T and B for top and bottom surface, respectively).
        \textbf{b}, Energy $E$ versus magnetic field $B$ plot and
        \textbf{c}, total electron density $n_\mathrm{tot}$ versus $B$ plot of Landau levels for $n_\mathrm{tot}=2.2\times10^{11}\ \mathrm{cm^{-2}}$ ($V^*_\mathrm{TG}\approx 0.1$~V), corresponding to the yellow line in Figure~5b (main text).
        \textbf{d}, $E$ versus $B$ plot and
        \textbf{e}, $n_\mathrm{tot}$ versus $B$ plot of Landau levels for $n_\mathrm{tot}=-1.8\times10^{11}\ \mathrm{cm^{-2}}$ ($V^*_\mathrm{TG}\approx-0.2$~V; black line in Figure~5b).
    }
  \label{fig:ext_kp}
\end{figure*}

In order to model the effect of the electrostatics on the band structure, we add a Hartree potential $U_\mathrm{H}(z)$ to the 8-orbital $k\cdot p$ Hamiltonian (for the latter, see e.g., Ref.~\cite{novik2005band}). The Hartree potential is a function of position $z$ in the growth direction only, and its functional dependence is determined by the carrier densities on the top and bottom surfaces. We construct it as a sum of two potential functions, $U_\mathrm{H,T}(z)$ and $U_\mathrm{H,B}(z)$, related to charges at the top and bottom surfaces, respectively, as illustrated in  Figure~\ref{fig:ext_kp}a. Assuming uniform charge densities within a distance $d_\mathrm{s}=8$~nm from the interfaces \cite{mahler2021vps}, $U_\mathrm{H,T}(z)$ and $U_\mathrm{H,B}(z)$ show a quadratic functional dependence near the top ($z=d_\mathrm{TI}/2$) and bottom ($z=-d_\mathrm{TI}/2$) interfaces, respectively, and vanish in the bulk. We assign the value of potential functions at the interfaces as $U_\mathrm{T} \equiv U_\mathrm{H,T}(d_\mathrm{TI}/2)$ and $U_\mathrm{B} \equiv U_\mathrm{H,B}(-d_\mathrm{TI}/2)$. Thus, $U_\mathrm{H}(z)$ is determined by three parameters: $U_\mathrm{T}$, $U_\mathrm{B}$, and the thickness of surface-state region $d_\mathrm{s}=8$ nm. The total carrier density $n_\mathrm{tot}$ relates to these parameters as
\begin{equation}
    n_\mathrm{tot}=\frac{\epsilon_\mathrm{r}\epsilon_\mathrm{0}}{ed_\mathrm{s}}(U_\mathrm{T}+U_\mathrm{B})
\label{eq:totden}
\end{equation}
where $\epsilon_\mathrm{r}$ is the dielectric constant at the surface state region \cite{mahler2021vps}.

In order to simulate the effect of a changing top gate, we only vary $U_\mathrm{T}$ and keep $U_\mathrm{B}$ constant.
This is justified by the screening of the gate electric field by the surface states \cite{brune2014dirac}.
For each given $U_\mathrm{H}$, we perform $k\cdot p$ calculations to obtain Landau levels as a function of magnetic field $B$ and energy $E$. For total density $n_\mathrm{tot}=2.2\times10^{11}\ \mathrm{cm^{-2}}$ and $n_\mathrm{tot}=-1.8\times10^{11}\ \mathrm{cm^{-2}}$ (corresponding to the densities shown in Figure~5 of the main text), we illustrate the Landau fan diagrams in Figures~\ref{fig:ext_kp}b and \ref{fig:ext_kp}d, respectively. (As input for the Hartree potential $U_\mathrm{H}$(z), we have used $U_\mathrm{T}=-18$ meV, $U_\mathrm{B}=-30$ meV, and $U_\mathrm{T}=+69$ meV, $U_\mathrm{B}=-30$ meV, respectively.)
For each Landau level dispersion $E(B)$, we calculate the corresponding carrier density $n_\mathrm{tot}(B)$, and thus map the energy-versus-field Landau fan to a density-versus-field fan, as shown for Figures~\ref{fig:ext_kp}c and \ref{fig:ext_kp}e for the aforementioned densities.

From the density-versus-field plot, we can readily extract the positions of the Landau levels by finding the intersections of the Landau level dispersion with constant density $n_\mathrm{tot}$, i.e., the yellow and black horizontal lines in Figures~\ref{fig:ext_kp}c and \ref{fig:ext_kp}e, respectively. (In the energy-versus-field diagrams, Figures~\ref{fig:ext_kp}b and \ref{fig:ext_kp}d, the Fermi energy as function of magnetic field is indicated as the yellow and black curves, respectively.)
The values where the intersection points occur determine the locations of the Landau transitions in the $\sigma_{xy}(B)$ dependence shown in Figures~5a and 5c of the main text.
By sweeping through intermediate density values $n_\mathrm{tot}$, we obtain Figure~5b of the main text. Some Landau levels may appear at more than one density value $n_\mathrm{tot}$. This remarkable behaviour occurs due to the intricate non-monotonic dependence of the Fermi level as function of $n_\mathrm{tot}$ which results from pinning  to the massive surface state \cite{mahler2021vps}.

\begin{figure*}[h]
  \includegraphics[width=120mm]{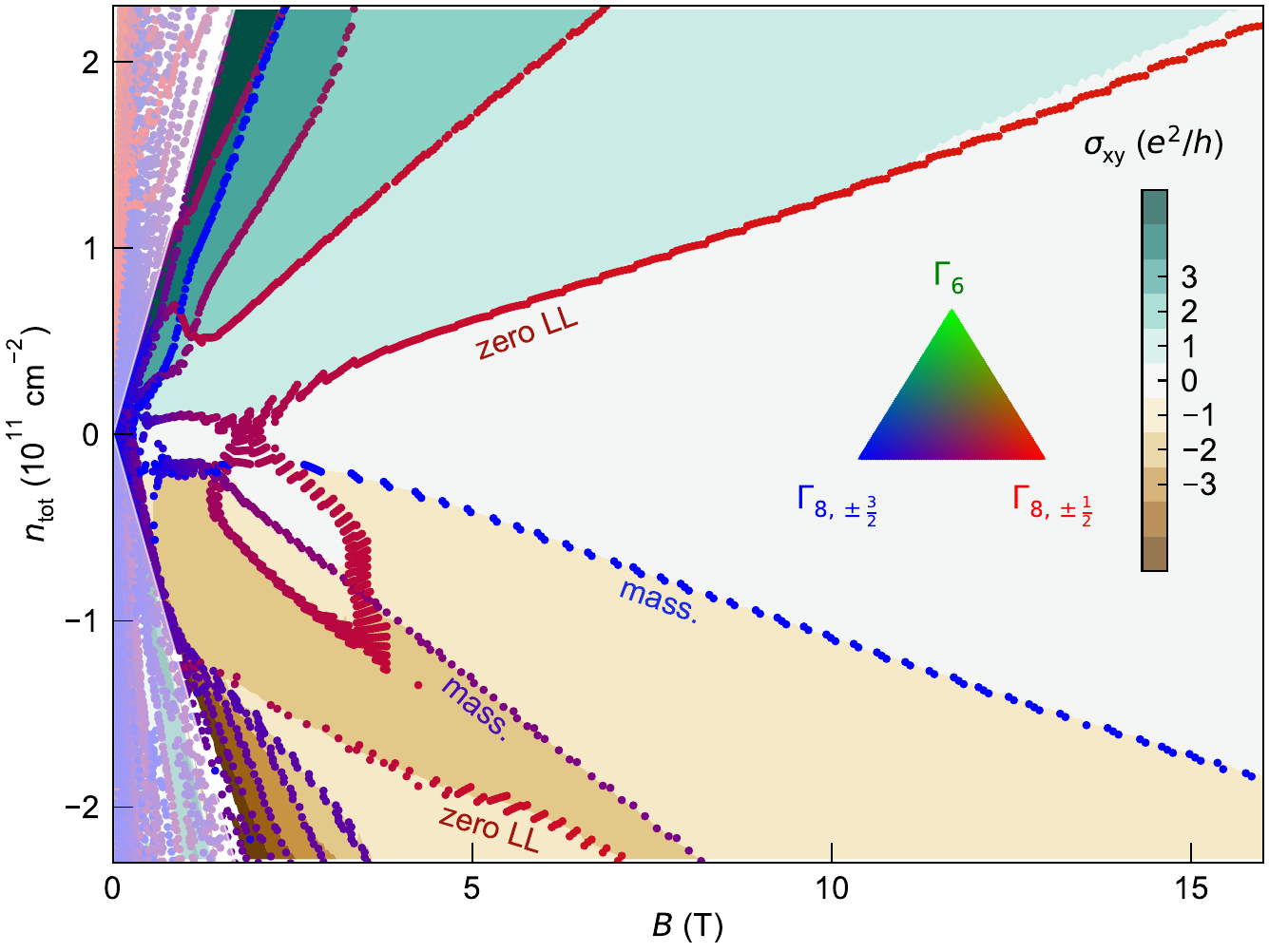}\\
  \caption{%
  		\textbf{Orbital character of calculated Landau level fan.} The displayed result is from the same calculations performed for Figure~5b (main text). The color code in the background is associated with the Hall conductivity $\sigma_{xy}$ in unit of $e^2/h$. The Landau levels are indicated by the dots, the color of which (legend in inset) indicates their orbital character (a combination of $\Gamma_{6}$, $\Gamma_{8,\pm1/2}$ and $\Gamma_{8,\pm3/2}$ orbitals). Note that the data at small fields and high $|n_\mathrm{tot}|$ (shaded region) is inaccurate due to limitations of the Landau level calculations.
    }
  \label{fig:ext_llfan}
\end{figure*}

Complementary to the location of the wave function (expectation value $\langle z\rangle$) indicated in Figure~5b (main text) as blue and red color, we also provide the orbital character of Landau levels in Figure~\ref{fig:ext_llfan}. The Landau level labelled ``mass.'' from the massive surface state originates from the $\Gamma_{8,\pm3/2}$ orbitals, and the ``zero LL'' Landau level is from the $\Gamma_{8,\pm1/2}$ orbitals, manifesting the band inversion at lower fields and its restoration to normal band ordering at higher fields.

\clearpage
\section{Supplementary Note 5: Theory of spectral asymmetry for a single surface state}
To analyze the contribution of a single surface state to $\sigma_{xy}$, we use an effective Hamiltonian in the basis $\{ \ket{\text{B};\uparrow}, \ket{\text{B};\downarrow} \}$ of the two spin components of the bottom surface state,
\begin{equation}
    H_\mathrm{B} = \mathcal{C} \sigma_0 + m_k \sigma_z - \mathcal{A}(k_y\sigma_x - k_x \sigma_y)
\end{equation}
where $\sigma_{x,y,z}$ is a set of Pauli matrices acting on the spin degree of freedom and $m_k = \mathcal{M} - \mathcal{B} |k|^2$ is an effective mass term (written up to second order in $|k|$) which needs to appear to cure the parity anomaly. In contrast to a single block of the BHZ model \cite{bhz2006}, the $\sigma_z$ terms of this Hamiltonian break time-reversal symmetry. The Dirac mass $m_k$ can only be non-zero in a finite magnetic field; for $B=0$, $\mathcal{M}=\mathcal{B}=0$. In our system, $\mathcal{M}$ is realized by a combination of a Zeeman field, the exchange coupling generated by the Mn doping, and the orbital field, that originates from the Peierls substitution. The quadratic $\mathcal{B}$ term arises from the coupling of the surface state with the bulk $j_z=\pm 3/2$ states in a magnetic field, and satisfies $\bhzcoeff{B}<0$.

We are interested in the evolution of the surface state in the presence of a magnetic field. Applying the Peierls substitution in the Landau gauge $\mathbf{k} \to \mathbf{k}+ e\mathbf{A}/\hbar$ with $\mathbf{A}=-y B \mathbf{e}_x$, we obtain the Landau level structure by replacing the canonical momentum operators with ladder operators \cite{Konig2008} 
\begin{align}\label{eq:LLn}
    E_{n\neq 0}^\pm &= \mathcal{C} + s\frac{\beta}{2} \pm \sqrt{n\alpha^2 + \left[\mathcal{M} - n \beta\right]^2} \\
    E_{n=0} &= \mathcal{C} - s\left[\mathcal{M} -\frac{\beta}{2} \right]
\end{align}
where $s=\sgn(eB)$ is the direction of the magnetic field, $\beta=2\mathcal{B}/l_B^2$, $\alpha=\sqrt{2}\mathcal{A}/l_B$, $l_B^2=\hbar/|eB|$, and $n$ is the Landau level index. The surface state forms a Landau level fan with a single decoupled $n=0$ Landau level. 

The single Landau level with $n = 0$ lies either above or below the charge neutrality point, creating an imbalance between the number of states above or below, i.e., a spectral asymmetry. Intuitively, the spectral asymmetry can be calculated by counting the Landau level states above and below the charge neutrality point $E_z$ \cite{NiemiSemenoff1983,bottcher2019survival},
\begin{equation}\label{eq:counting}
    \eta_B = \sum_{E>E_z} 1 - \sum_{E<E_z} 1.
\end{equation}
Here, the charge neutrality point is at $E_z = \mathcal{C}$, which is determined by demanding a vanishing particle number in the ground state \cite{Bottcher2020}.
In view of the divergent sums of Equation~\eqref{eq:counting}, we apply a heat-kernel regularization following Ref.~\cite{bottcher2019survival} in order to make $\eta_B$ well-defined.

By definition, $\eta_B$ only changes when a Landau level crosses the charge neutrality point as one varies the magnetic field.
From Equation~\eqref{eq:LLn}, one can see that the $n\neq0$ Landau levels never do so. Only the $n=0$ state can cross $E_z$, i.e., whenever $B=s \hbar \mathcal{M}/(e \mathcal{B})$ is satisfied. This results in a jump of $\pm 2$ in the spectral asymmetry.
We apply the method used in Ref.~\cite{bottcher2019survival} and obtain 
\begin{equation}
    \eta_B = - s \left[\mathrm{sgn}(\mathcal{B}) + \sgn\left(\mathcal{M} - \frac{\mathcal{B}}{l_B^2}\right)\right].
\end{equation}
The contribution $\sigma_{xy}^{\mathrm{BSS}}$ of the bottom surface state to the Hall conductance at charge neutrality can be evaluated as 
\begin{equation}
 \label{eq:sxyBSSmethods}
 \sigma_{xy}^{\mathrm{B}}
 =  - \frac{e^2}{2h} s \left[\mathrm{sgn}(\mathcal{B}) + \sgn\left(\mathcal{M} - \frac{\mathcal{B}}{l_B^2}\right)\right]
\end{equation}
from a calculation of the Chern number for $H_\mathrm{B}$. This contribution 
relates to the spectral asymmetry as $\sigma_{xy}^{\mathrm{B}} = (e^2/h)(\eta_B/2)$, and can take discrete values depending on the competition between $\mathcal{M}$ and $\mathcal{B}/l_B^{2}$. If the signs of both terms in Equation~\eqref{eq:sxyBSSmethods} are equal, we find a finite contribution of $\pm e^2/h$ to the total $\sigma_{xy}$ due to the parity anomaly. If the terms have opposite sign, the contribution vanishes.

The above is a minimal model with two degrees of freedom, to illustrate the effect of the spectral asymmetry on $\sigma_{xy}$. The principle remains valid for more realistic band structures with many more degrees of freedom taken into account, e.g., the bulk valence band states and the massive surface states at the opposite surface. In the main text, we have retrieved the signature of spectral asymmetry in a full $k\cdot p$ model.

\end{document}